\documentclass[prl,twocolumn,amsmath,amssymb,notitlepage]{revtex4-1}

\usepackage[utf8]{inputenc}
\usepackage{hyperref,braket,graphicx,color}
\DeclareMathOperator{\sech}{sech}

\graphicspath{{imagesSQD/}}

\begin{document}
\title{Minimum Disturbance Rewards with Maximum Possible Classical Correlations}

\author{Varad R. Pande}
\affiliation{Department of Physics, Indian Institute of Science Education and Research Pune, India 411008}
\email{varad\_pande@yahoo.in}
\author{Anil Shaji}
\affiliation{School of Physics, Indian Institute of Science Education and Research Thiruvananthapuram, India 695016}
\date{\today}

\begin{abstract}
	Weak measurements done on a subsystem of a bipartite system having both classical and nonClassical correlations between its components can potentially reveal information about the other subsystem with minimal disturbance to the overall state. We use weak quantum discord and the fidelity between the initial bipartite state and the state after measurement to construct a cost function that accounts for both the amount of information revealed about the other system as well as the disturbance to the overall state. We investigate the behaviour of the cost function for families of two qubit states and show that there is an optimal choice that can be made for the strength of the weak measurement.    
\end{abstract}

\maketitle


NonClassical correlations in quantum states including, but not limited to, entanglement has been a topic of significant interest in the recent past because of the potential and promise held forth by quantum information processing and quantum technologies~\cite{Horodecki:2009gb,modi2012classical,nielsen2010quantum}. Ollivier and Zurek~\cite{zurek2003quantum} and independently, Henderson and Vedral~\cite{Henderson:JournalOfPhysicsAMathematicalAndGeneral:2001}, noted that mixed quantum states allowed for the possibility of having nonClassical correlations other than entanglement and quantified the same in terms of the quantum discord. A variety of alternate measures of nonClassical correlations in a bipartite quantum state were subsequently proposed~\cite{luo08a,modi2012classical,wu09a}. A general strategy followed in constructing measures of nonClassical correlations is to subtract the `classical' correlations in a bipartite state from the `total' correlations; treating what remains as a quantifier of the nonClassical or quantum correlations in the state~\cite{LANG:IntJQuanumInform:2011}. 

Typically, entropic measures like the mutual information and relative entropy are used to quantify the correlations in constructing the various measures. Quantifying the total correlations in a bipartite quantum state is straightforward, for instance, using the quantum mutual information. However, defining the `classical' part of the total correlations is often a relatively ambiguous task. One strategy is to posit classical observers measuring one or both of the subsystems so as to quantify the correlations in the resultant measurement statistics. To achieve this, the classical observers utilize the classical counterpart of the same entropic measure of quantum correlations that was used to quantify the total correlations. Significantly though, in the quantum case, the measurement statistics depend on the measurement done. This necessitates a further maximisation of the measure of classical correlations over all measurement strategies in order to disambiguate the discord-like measure to the maximum extent possible. In the ensuing treatment, quantum discord is considered as the example of nonClassical correlations. The total correlations in a bipartite state $\rho_{AB}$ are measured in terms of the quantum mutual information defined as 
\begin{equation}
	\label{eq:mututalI}
	I(A:B)=S(\rho_{A})+S(\rho_{B})-S(\rho_{AB}), 
\end{equation}
where $S(\rho) = - {\rm tr}[\rho \log \rho]$ is the vonNeumann entropy of a state $\rho$ and $\rho_{A,B} = {\rm tr}_{B,A} (\rho_{AB})$ are the reduced (partial trace) density matrices of subsystems $A$ and $B$. Based on a general (POVM) measurement on subsystem $B$ given by $\{E^{B}_{j}\}$ and the resultant measurement statistics $\{p^{B}_{j}\}$ we can define the `classical' mutual information between $A$ and $B$ as
\begin{equation}
	\label{eq:mutualJ}
	J(A:B) = S(\rho_{A}) - S(A|B), 
\end{equation}
where 
\[ S(A|B) = \sum_{j} p^{B}_{j} S(\rho_{A|E^{B}_{j}}),\] 
is the conditional entropy of subsystem $A$ conditioned on the measurement on $B$. Here, $\rho_{A|E^{B}_{j}}$ is the post-measurement state of $A$ corresponding to the result labeled by $j$ obtained on measuring $B$. Quantum discord is defined as
\begin{equation}
	\label{eq:discord}
	{\cal D}(A,B) \equiv I(A:B) - \max_{\{E^{B}_{j} \}} J(A:B). 
\end{equation}
Note that $I(A:B) = J(A:B)$  and ${\cal D} = 0$ as a consequence of Bayes' theorem if the quantum state $\rho_{AB}$ is replaced by a joint probability distribution $p(A,B)$ describing a bipartite classical system. 

Any discussion of a measurement on a quantum system is incomplete without the unavoidable disturbance it causes on the system. In fact, the original context in which Ollivier and Zurek introduced quantum discord is by discussing the disturbance caused on one subsystem of a bipartite state due to projective measurements performed on the other. While this aspect was rarely considered in subsequent discussions on quantum discord and other measures of nonClassical correlations, the question was brought back into focus recently by exploring the behaviour of discord and discord-like measures when the measurements on one or both subsystems were restricted to weak quantum measurements~\cite{singh2014quantum,shaji2015weak}. The weak measurement formalism proposed by Aharonov, Albert and Vaidman~\cite{aharonov1988result}, and elucidated further in \cite{duck1989sense}, gave a means of quantifying the disturbance on a quantum state due to the interaction of the system with the `pointer' of a measuring device. Recently, there has been progress in investigating weak measurements and their interesting consequences including weak value amplification in the laboratory as well~\cite{dixon2009ultrasensitive,hosten2008observation,Putz:2016tm}. Oreshkov and Brun~\cite{oreshkov2005weak} recast weak measurement using the language of POVMs and further showed that any generalised measurement can be modeled as a sequence of weak measurements. In the following we take the approach in~\cite{oreshkov2005weak} and use a POVM to model weak measurements because our primary focus is on the limited changes to the measured system due to the weak measurement and we do not consider here the post-selection through projective measurements that is a part of the approach in~\cite{aharonov1988result}.

Steering clear of the foundational issues raised by weak measurements including those related to complex weak values, weak value amplification~\cite{dressel2012significance,knee2016weak} etc, we motivate the investigations in this Letter through broad considerations of quantum information processing. The input information entered on to suitable quantum registers in a quantum information processing protocol is typically manipulated by introducing additional computational space in the form of registers of ancilla qubits. Readout of the output also often involves ancilla registers depending on the measurement model employed. NonClassical correlations including entanglement that get generated between the registers and all the quantum bits in them is recognised as a resource that, under the right circumstances, enables the quantum information processor to perform its task exponentially faster than equivalent classical entities. Readout of the information content in the quantum registers as classical, human readable, information at intermediate or final stages of the information processing protocol is of interest to us in the following because such steps entail measurements typically on some of the registers involved in the computation. One can ask the question whether these measurements can be made gently enough in a manner that while revealing the classical information output that is desired, they preserve the quantum resources including nonClassical correlations between the registers to the maximum extent possible so that these resources may be used again. 

We introduce a cost function that quantifies both the extent to which the measurements done on one subsystem can reveal information residing on the other subsystem using the notion of weak discord~\cite{singh2014quantum,shaji2015weak}, as well as the disturbance to the overall state due to the (weak) measurement on the subsystem. Note that the cost function is defined only in the bipartite context which appears frequently in information processing protocols where we have an ancilla register which is read-out and a memory register that holds the processed quantum information. Minimising the cost function would mean optimal extraction of the desired classical information from the quantum registers of the information processor with minimal disturbance to its state. 

To quantify the extent to which weak measurements on one subsystem can reveal information about the other due to the {\em classical} correlations that exist between the two, we start with the weak quantum discord. Note that the quantity we refer to as weak quantum discord following~\cite{shaji2015weak} is called super quantum discord in~\cite{singh2014quantum} and the difference in points of view that leads to two names that seemingly convey opposite meanings is discussed in detail in~\cite{shaji2015weak}. In what follows, we restrict our discussion to a bipartite quantum system with two qubits even though it can be easily generalised to two registers of qubits. As in~\cite{singh2014quantum}, we express the non-projective measurements that preserve the subsystem $B$ of a quantum system $AB$ to the desired extent even after the act of measurement in terms of a two outcome POVM~\cite{oreshkov2005weak} with elements:
\begin{eqnarray}
	\label{eq:measureOp}
	P_{x} & = & \sqrt{\frac{1-\tanh(x)}{2}}\Pi_0 + \sqrt{\frac{1+\tanh(x)}{2}}\Pi_1, \nonumber \\
	P_{-x}& = & \sqrt{\frac{1+\tanh(x)}{2}}\Pi_0 + \sqrt{\frac{1-\tanh(x)}{2}}\Pi_1,
\end{eqnarray}
where $x$ is a parameter that denotes the strength of the measurement process and $\Pi_0$ and $\Pi_1$ are two
orthogonal projectors forming a complete set such that $\Pi_0 + \Pi_1 = \openone$. After the measurement, the normalised post measurement state of subsystem $A$ is given by:
	\begin{equation}
	\rho_{A|P^{B}_{\pm x}} = \frac{{\rm Tr}_B[(\openone \otimes P^{B}_{\pm x})\rho_{AB}(\openone \otimes P^{B}_{\pm x})]}{{\rm Tr}_{AB}[(\openone \otimes P^{B}_{\pm x})\rho_{AB}(\openone \otimes P^{B}_{\pm x})]}\label{eq10}
	\end{equation}
with respective probabilities 
\begin{equation}
p_{w}(\pm x) = {\rm Tr}_{AB}[(\openone \otimes P^{B}_{\pm x})\rho_{AB}(\openone \otimes P^{B}_{\pm x})]\label{eq11}.
\end{equation}
The subscript $w$ indicates that the probabilities arise from weak measurements on subsystem $B$. In what follows, this subscript is used for quantities computed from the results of the weak measurements and the same symbols without the subscript denotes quantities computed from the results of normal projective measurements. The conditional entropy for subsystem $A$ conditioned on the measurements on $B$ is then
\[ S_w(A|B)=p_{w}(x)S_w(\rho_{A|P^{B}_{x}})+p_{w}(-x)S_w(\rho_{A|P^{B}_{-x}}).\]
Like in the case of ordinary quantum discord in \eqref{eq:discord}, we can now define the `classical' mutual information as
\[ J_{w}(A:B) = S(\rho_{A}) - S_{w}(A|B) \]
and the  weak quantum discord as:
\begin{equation}
\label{eq12}
{\cal D}_{w}(A,B) := I(A:B) - \max\limits_{\{\Pi_{j}^{B}\}}J_w(A:B).
\end{equation} 
The maximisation here is limited to one over all sets of projectors $\Pi_{j}^{B}$ and not over the parameter $x$ corresponding to the strength of the measurement. For large values of $x$, $\tanh x \rightarrow 1$ and weak discord reduces to normal discord since $P_{x}$ and $P_{-x}$ become a pair of orthogonal projectors. 

Since weak measurements on subsystem $B$ reveal less about $A$, the conditional entropy $S_w (A|B)$ is greater than $S(A|B)$. This means that the weak quantum discord is greater than the normal discord. We can therefore characterise how well (or how badly) the weak measurements leverage the classical correlations that may exist between subsystem $A$ and $B$ to reveal information about $A$ upon measuring $B$ by considering the quantity, 
\begin{equation}
	\label{eq:disdiff}
	\Delta {\cal D} = {\cal D}_{w}(A,B) - {\cal D}(A,B).
\end{equation}
This quantity will be large when the weak measurements on $B$ reveal very little information on $A$ because then weak quantum discord would essentially count all the correlations in the bipartite state, classical or otherwise while normal discord, by construction, would count only the quantum correlations.  

The advantage one gets by doing weak measurements on $B$ is that the disturbance on the state of $AB$ is kept within well defined limits. We quantify the disturbance on the state of $AB$ in terms of the Fidelity of the state of $\tilde{\rho}_{AB}$ of the two qubit system after the measurement of $B$ with the state $\rho_{AB}$ of the system before the measurement. In particular, as the figure of merit of the disturbance we use
\begin{equation}
	\label{eq:fiddiff}
	\Delta F = 1-F(\rho_{AB}, \tilde{\rho}_{AB}), 
\end{equation}
where 
\begin{equation}
	\label{eq13}
	F(\rho,\sigma)={\rm Tr} \Big[\sqrt{\sqrt{\rho}\sigma\sqrt{\rho}} \Big],
\end{equation}
and
\begin{eqnarray}
	\label{eq:postmeasure}
	 \tilde{\rho}_{AB} & = &  p_{w}(x) (\openone \otimes P_{x}^{B}) \rho_{AB} (\openone \otimes P_{x}^{B}) \nonumber \\
	 && \qquad + \, p_{w}(-x) (\openone \otimes P_{-x}^{B}) \rho_{AB} (\openone \otimes P_{-x}^{B}). 
\end{eqnarray}
Combining the two desirable features -- revealing as much as possible about subsystem $A$ by measuring $B$ and simultaneously disturbing the state of $AB$ to the minimum extent possible, we define a cost function:
\begin{eqnarray}
	\label{eq:cost}
	{\cal C} & = &  \Delta F + \Delta {\cal D}  \nonumber \\
	& = & 1-F(\rho_{AB}, \tilde{\rho}_{AB}) +{\cal D}_{w}(A,B) - {\cal D}(A,B)
\end{eqnarray}
We explore the behaviour of the cost function ${\cal C}$ below for several families of two qubit states. For computing the fidelity and the weak discord, it is useful to note that the measurement operators in Eq.~(\ref{eq:measureOp}) induce the following transformation on the Bloch vector representing the state of qubit $B$:
\[ (r_x,r_y,r_z)\rightarrow (r_x \sech(x),r_y\sech(x),r_z), \]
which, in turn, corresponds to the phase damping channel.  Note that we have chosen the positive $z$ axis of the Bloch sphere to be oriented along the direction of $\Pi_{0}$ in obtaining the transformation above.

\begin{figure}[!htb]
	\resizebox{8.5cm}{5.5cm}{\includegraphics{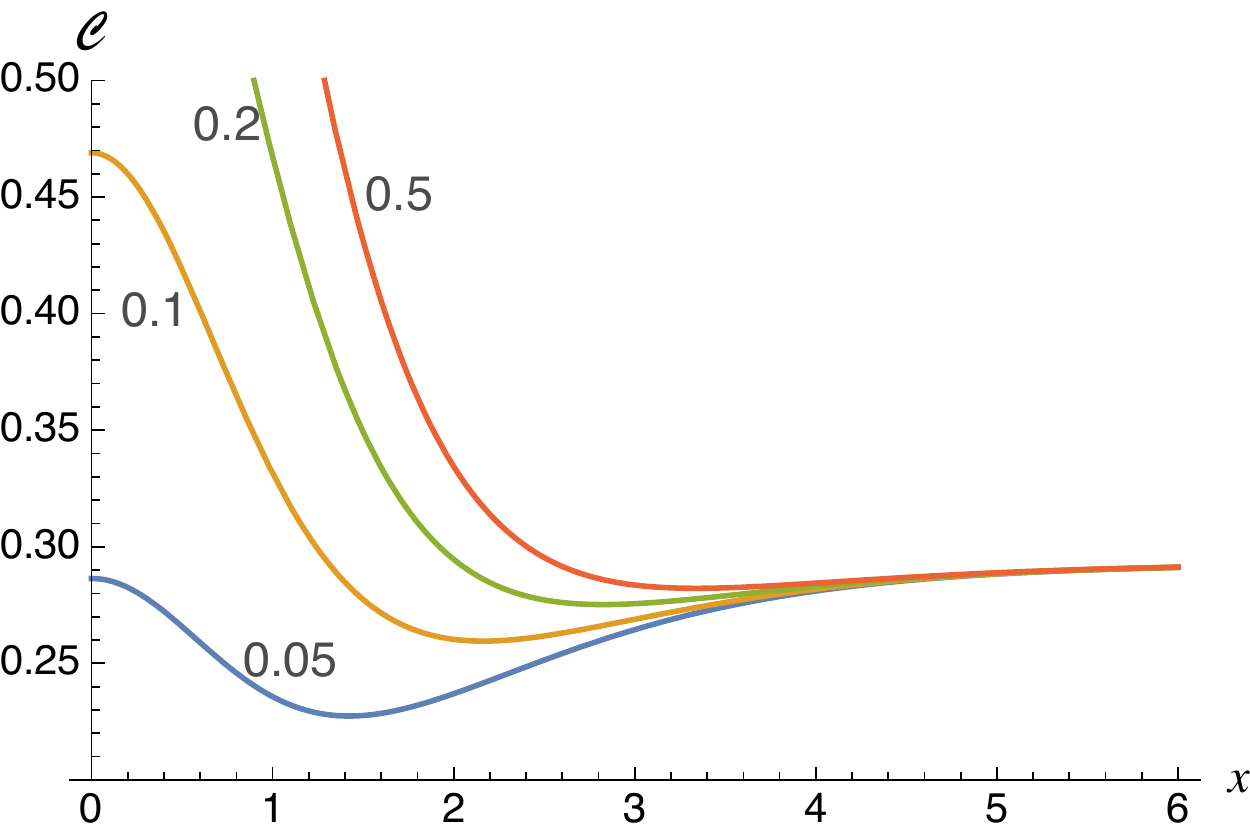}}
	\caption{The cost function ${\cal C}$ as a function of $x$ for the entangled state in Eq.~(\ref{eq:entstate1}) corresponding to $\lambda_{0} = 0.05$, $0.1$, $0.2$ and $0.5$ as labeled in the plot. We see that the cost function has a minimum in all these cases indicating that a trade off between disturbance to the state and revealing of information about subsystem $A$ through measurements on subsystem $B$ is possible. \label{fig:entstate}}
\end{figure}

\begin{figure}[!htb]
	\resizebox{8.5cm}{5.5cm}{\includegraphics{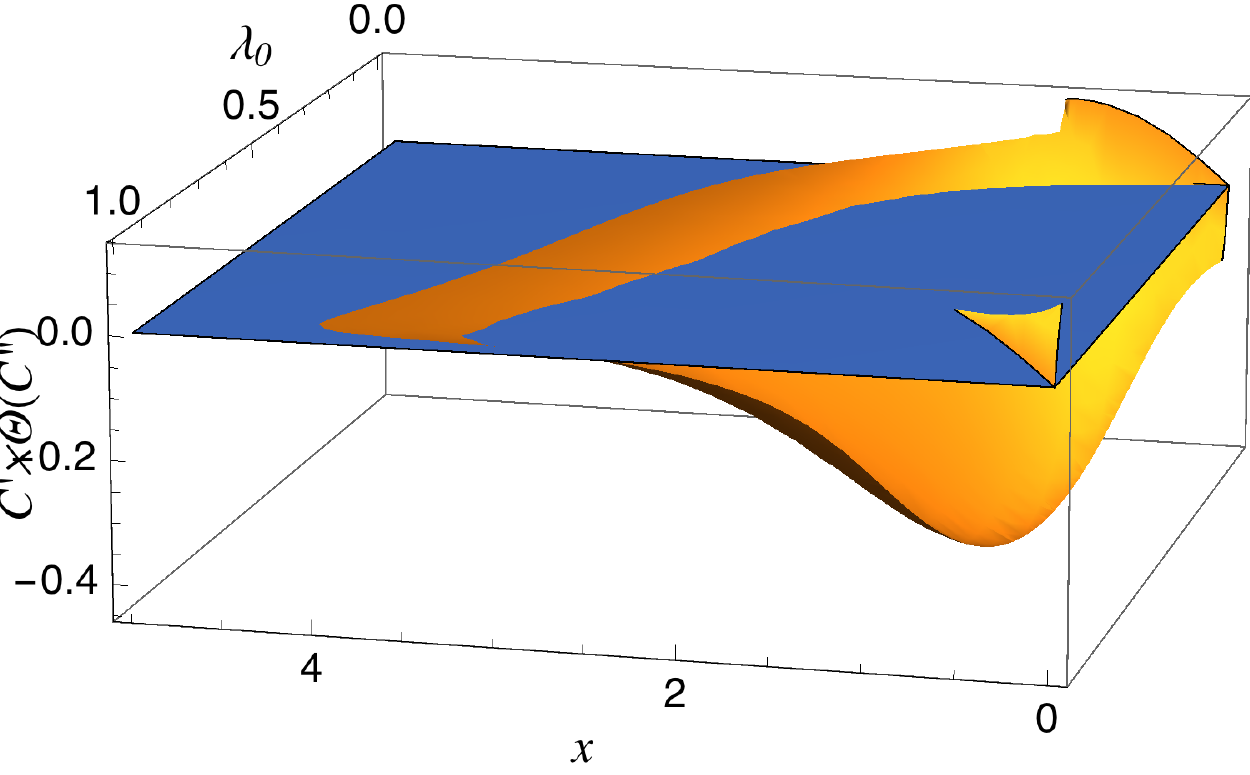}}
	\caption{The first derivative cost function ${\cal C}'$ is plotted as a function of $x$ and $\lambda_{0}$ in the regions where the second derivative ${\cal C}''$ is positive. The ${\cal C}'=0$ plane (blue) is also shown. We see that ${\cal C}'$ has only one zero crossing as a function of $x$ for each value of $\lambda_{0}$ with ${\cal C}'' >0$ indicating that the cost function ${\cal C}$ has only one minimum with respect to $x$}. \label{fig:singlemin}
\end{figure}

A two qubit pure entangled state furnishes a tractable and simple first example that we work out. The state has a Schmidt decomposition of the form:
\begin{equation}
	\label{eq:entstate1}
	{\ket{\psi}}_{AB}=\sqrt{\lambda_0}\ket{00}+\sqrt{\lambda_1}\ket{11},
\end{equation}
where $\lambda_0$ and $\lambda_1$ are the Schmidt coefficients. Since ${\ket{\psi}}_{AB}$ is a pure state, the discord is equal to the vonNeumann entropy of either one of the two subsystems, 
\begin{equation}
	\label{eq:disc2}
	{\cal D}(A,B)=-\lambda_0\log_{2}(\lambda_0)-\lambda_1\log_{2}(\lambda_1)
\end{equation}
The weak measurement operators in an arbitrary measurement basis oriented in the direction $(\theta,\phi)$ in the Bloch sphere are:
\begin{equation}
	\label{eq16}
	P^{B}_{\pm x}=\sqrt{\frac{1\mp\tanh(x)}{2}} \Pi_{\psi}+\sqrt{\frac{1\pm\tanh(x)}{2}}  \Pi_{\bar{\psi}}
\end{equation}
where $\{\ket{\psi(\theta,\phi)},\ket{\bar{\psi}(\theta,\phi)}\}$ is the arbitrary single qubit basis. The weak discord for ${\ket{\psi}}_{AB}$ is obtained as~\cite{singh2014quantum}:
\begin{eqnarray}
	\label{eq:wdisc2}
{\cal D}_w(A,B) & = & -\lambda_0\log_{2}\lambda_0-\lambda_1\log_{2}\lambda_1 \nonumber \\
&& \quad - \,\min_{\theta} \Big\{ \sum_{y=\pm x}p_{w}(y) \big[ k_{+}(y)\log_{2} k_{+}(y)  \nonumber \\
&& \qquad \qquad + \, k_{-}(y)\log_{2} k_{-}(y) \big] \Big\}
\end{eqnarray}
where 
\[ k_{\pm}(y)=\frac{1}{2}\bigg[1\pm\sqrt{1-\frac{\lambda_0\lambda_1}{p_{w}(y)^2\cosh^2(y)}} \bigg], \]
with 
\[ p_{w}(\pm x) = \frac{1}{2} \big[ 1 \pm (\lambda_{1} - \lambda_{0}) \cos (\theta) \tanh (x)  \big]. \]
Note that, since $\tanh(x)$ is an odd function of $x$, $p_{w}(-x)=p_{w}(x)$. Since both $p_{w}(x)$ and $\cosh(x)$ are even functions of $x$, it follows that both $k_{+}$ and $k_{-}$ are also even functions. By computing the post measurement state of the two qubits $\tilde{\rho}_{AB}$ as in Eq.~(\ref{eq:postmeasure}), we can compute the fidelity to be 
\begin{eqnarray}
	\label{eq:fid2}
	F & = & \frac{1}{2} \Big[ 2 \left(\lambda_0^2+\lambda_1^2\right)-\cos (2 \theta ) (\lambda_0-\lambda_1)^2 (\text{sech}(x)-1) \nonumber \\
&& \qquad \qquad + \, (4 \lambda_0 \lambda_1+1) \text{sech}(x)+1 \Big]^{1/2}
\end{eqnarray} 
From Eqs.~(\ref{eq:disc2}) and (\ref{eq:wdisc2}) we find 
\begin{eqnarray}
	\label{eq:discdiff2}
	\Delta {\cal D} & = & - \,\min_{\theta} \Big\{ \sum_{y=\pm x}p_{w}(y) \big[ k_{+}(y)\log_{2} k_{+}(y)  \nonumber \\
&&  \qquad \qquad + \, k_{-}(y)\log_{2} k_{-}(y) \big] \Big\}.
\end{eqnarray}
Using $k_{+}(x) = k_{+}(-x)$ and $k_{-}(x) = k_{-}(-x)$ we find that the minimum in the above equation is reached for $\theta = \pi/2$. For the same value of $\theta$ we have a maximum for the fidelity $F$ in Eq.~(\ref{eq:fid2}) as well. In Fig.~\ref{fig:entstate} the cost function ${\cal C}$ is plotted as a function of $x$ for the entangled state in Eq.~(\ref{eq:entstate1}). We see that the cost function has a minimum in each case and the minimum gets progressively shallow as $\lambda_{0}$ approaches $\lambda_{1}$. We expect at most one minimum for the cost function $\cal{C}$ because out of its two components, $\Delta F$ is a monotonically and smoothly increasing function of the disturbance due to the measurement, parameterised by $x$, while $\Delta {\cal D}$ is a monotonically decreasing function of $x$. This behaviour expected from the physical meanings of $\Delta {\cal D}$ and $\Delta F$ is seen to be true from Fig.~\ref{fig:singlemin} where the derivative ${\cal C}'$ of ${\cal C}$ with respect to $x$ is plotted as a function of both $x$ and $\lambda_{0}$ (with $\lambda_{1} = 1-\lambda_{0}$ and $\theta=\pi/2$). In the plot ${\cal C}'$ is multiplied by $\Theta({\cal C}'')$ where $\Theta$ is the Heaviside unit step function so that the first derivative is plotted only in the region where the second derivative is positive. We see that the first derivative has only one zero crossing as a function of $x$ across the ${\cal C}'=0$ plane  for each value of  $\lambda_{0}$ and positive ${\cal C}''$ indicating a single minimum for ${\cal C}$. The existence of the minimum in each case indicates that it is possible to set a value for the measurement strength $x$ such that optimal extraction of information about subsystem $A$ through measurements on $B$ with minimal disturbance of the overall state is achieved. When $\lambda_{0} = \lambda_{1}$ we have the Bell state $(|00\rangle + |11\rangle/\sqrt{2})$. As we approach the Bell state, we see that the minimum of ${\cal C}$ is very shallow and close to the saturation value for large $x$ indicating that weak measurements offer little advantage. This is to be expected for the Bell state knowing that the state of $B$ gives no information about the state of $A$ since both reduced states are fully mixed.  

\begin{figure}[!htb]
	\resizebox{8.5cm}{5.5cm}{\includegraphics{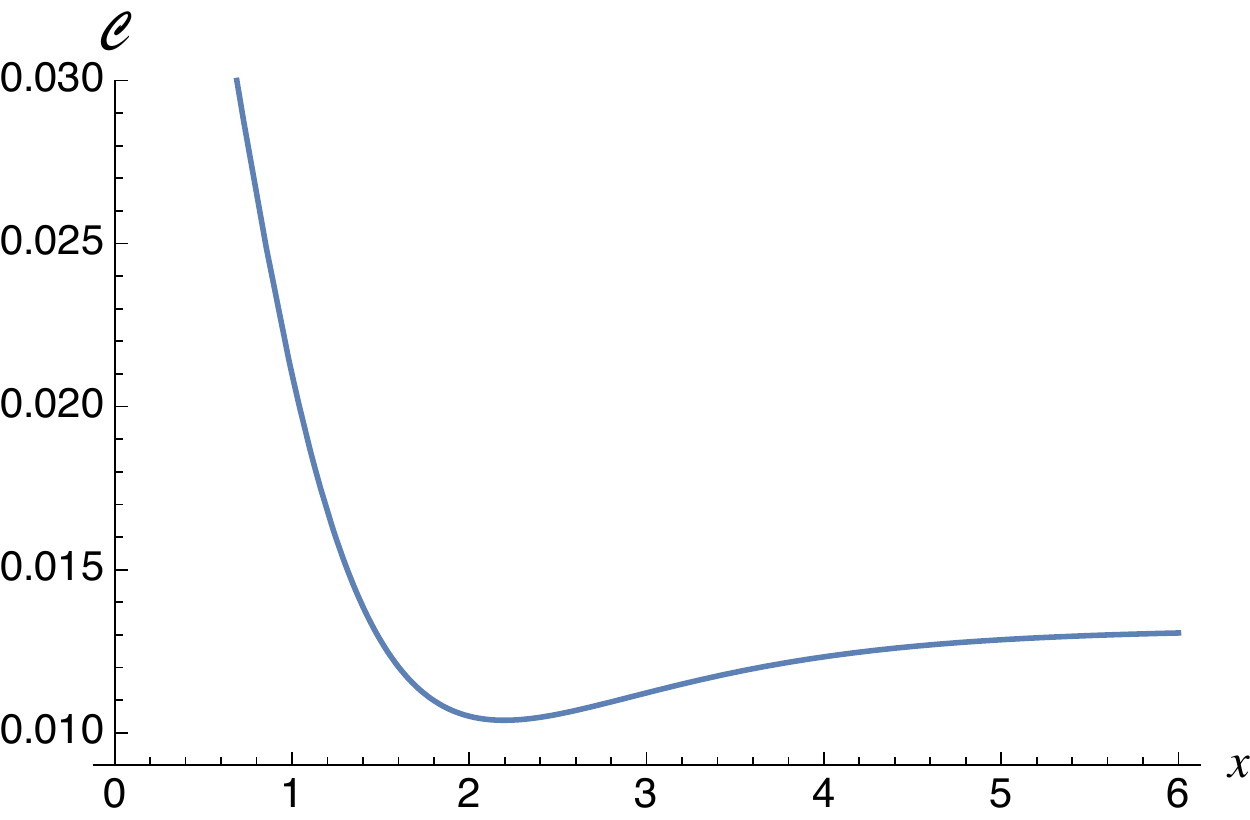}}
	\caption{The cost function ${\cal C}$ as a function of $x$ for the Werner state corresponding to $z = 0.25$. We see that the cost function has a minimum indicating that a trade off between disturbance to the state and revealing of information about subsystem $A$ through measurements on subsystem $B$ is possible. \label{fig:wernerstate}}
\end{figure}

As an example of a mixed quantum state for which the cost function can be computed and explored, we take the Werner state:
\begin{equation}
	\label{werner}
	\rho_{AB}=z\ket{\psi^-}\bra{\psi^-}+\frac{(1-z)}{4}I,
\end{equation}
the weak quantum discord for the Werner state is,
\begin{eqnarray}
	\label{eq23}
	{\cal  D}_w(A,B) \! & = & \! 1 + 3 \frac{(1-z)}{4}\log_{2}\frac{1-z}{4} \nonumber \\
	&&  + \, \frac{1+3z}{4}\log_{2}\frac{1+3z}{4} \nonumber \\
	&&  -   \frac{1-z\tanh(-x)}{2} \log_{2} \frac{1-z\tanh(-x)}{2}  \nonumber \\
	&&  -   \frac{1-z\tanh(x)}{2} \log_{2} \frac{1-z\tanh(x)}{2}. 
\end{eqnarray}
The expression above reduces to the discord~\cite{guo2014quantum} when $x \rightarrow \infty$.  The fidelity between initial state and the two qubit post measurement state is given by a very long expression which is not reproduced here. Using the weak discord and fidelity, the cost function $C$ can be computed and it is plotted  in Fig.~{\ref{fig:wernerstate}}. Again, we see that there is an optimal choice of the measurement strength corresponding to the best trade off between revealing the classical correlations present in the state and the disturbance to the overall state. 

\begin{figure}[!htb]
	\centering
	\resizebox{8.5cm}{5.5cm}{\includegraphics{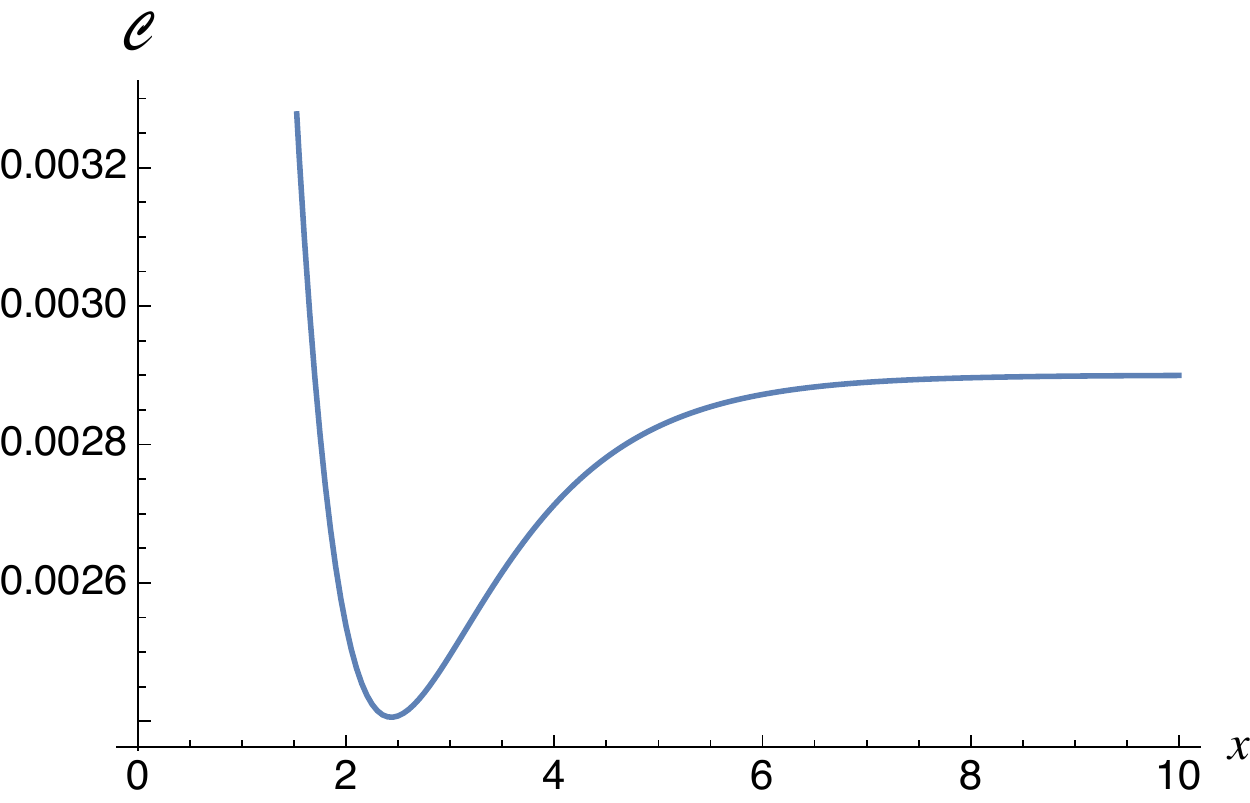}}
	\caption{The cost function ${\cal C}$ as a function of $x$ for a general two qubit state corresponding to $a_1=0.01, a_2=0.1, a_3=0.22, b_1=0.1, b_2=0.03, b_3=0.5, c_1=0.1, c_2=0.02, c_3=0.2$. }
	\label{fig 4}
\end{figure}

We can further generalise to an arbitrary two qubit state (upto a unitary equivalence~\cite{luo2008quantum}) which can be written in the form:
\begin{equation}
	\label{eq:arbState}
	\rho_{AB}=\frac{1}{4}[I+(\vec{a}.\vec{\sigma}\otimes \openone)+(\openone \otimes \vec{b}.\vec{\sigma})+\sum_{i=1}^{3}c_i(\sigma_i\otimes\sigma_i)]
\end{equation}
In this case also, we find that the cost function ${\cal C}$ has the same behaviour as in the previous examples and it shows a minimum as in Fig.~\ref{fig 4}. By doing an analysis similar to that done for the two qubit pure state, from numerical computation of the first and second derivatives of the cost function, it can be shown that it has at most one minimum for both examples of mixed states considered above.

To sum up, we have shown the existence of a single minimum for the cost function ${\cal C}$ at a particular value of the measurement strength $x$ for several families of two qubit states and also elaborated on the importance of having such a minimum. It enables us to extract maximum possible classical correlations from a bipartite quantum state with the least disturbance possible. Extension of weak discord to higher dimensional quantum states following such extensions for normal Discord\cite{vinjanampathy2011calculation,yurischev2015quantum,buscemi2013time,guo2016pairwise,rulli2011global} and establishing an operational interpretation~\cite{madhok2011interpreting,cavalcanti2011operational} for the cost function remains to be done.  The notion of weak quantum discord opens up new possibilities in quantum information processing for development of quantum information processing protocols which value the non-disturbance of quantum states~\cite{dakic2012quantum,pirandola2014quantum}.  

{\bfseries Acknowledgements} V.~R.~P.~acknowledges the support of DST-INSPIRE, IISER Pune and IISER Thiruvananthapuram for giving him the opportunity of doing summer projects which enabled this collaboration. A.~S.~acknowledges the support from the FAST grant for the Centre for Computation, Modelling and Simulation at IISER TVM from the Ministry for Human Resources Development, Government of India.

\bibliography{SQDoptquantdraft}

\begin{thebibliography}{29}%
\makeatletter
\providecommand \@ifxundefined [1]{%
 \@ifx{#1\undefined}
}%
\providecommand \@ifnum [1]{%
 \ifnum #1\expandafter \@firstoftwo
 \else \expandafter \@secondoftwo
 \fi
}%
\providecommand \@ifx [1]{%
 \ifx #1\expandafter \@firstoftwo
 \else \expandafter \@secondoftwo
 \fi
}%
\providecommand \natexlab [1]{#1}%
\providecommand \enquote  [1]{``#1''}%
\providecommand \bibnamefont  [1]{#1}%
\providecommand \bibfnamefont [1]{#1}%
\providecommand \citenamefont [1]{#1}%
\providecommand \href@noop [0]{\@secondoftwo}%
\providecommand \href [0]{\begingroup \@sanitize@url \@href}%
\providecommand \@href[1]{\@@startlink{#1}\@@href}%
\providecommand \@@href[1]{\endgroup#1\@@endlink}%
\providecommand \@sanitize@url [0]{\catcode `\\12\catcode `\$12\catcode
  `\&12\catcode `\#12\catcode `\^12\catcode `\_12\catcode `\%12\relax}%
\providecommand \@@startlink[1]{}%
\providecommand \@@endlink[0]{}%
\providecommand \url  [0]{\begingroup\@sanitize@url \@url }%
\providecommand \@url [1]{\endgroup\@href {#1}{\urlprefix }}%
\providecommand \urlprefix  [0]{URL }%
\providecommand \Eprint [0]{\href }%
\providecommand \doibase [0]{http://dx.doi.org/}%
\providecommand \selectlanguage [0]{\@gobble}%
\providecommand \bibinfo  [0]{\@secondoftwo}%
\providecommand \bibfield  [0]{\@secondoftwo}%
\providecommand \translation [1]{[#1]}%
\providecommand \BibitemOpen [0]{}%
\providecommand \bibitemStop [0]{}%
\providecommand \bibitemNoStop [0]{.\EOS\space}%
\providecommand \EOS [0]{\spacefactor3000\relax}%
\providecommand \BibitemShut  [1]{\csname bibitem#1\endcsname}%
\let\auto@bib@innerbib\@empty
\bibitem [{\citenamefont {Horodecki}\ \emph {et~al.}(2009)\citenamefont
  {Horodecki}, \citenamefont {Horodecki}, \citenamefont {Horodecki},\ and\
  \citenamefont {Horodecki}}]{Horodecki:2009gb}%
  \BibitemOpen
  \bibfield  {author} {\bibinfo {author} {\bibfnamefont {R.}~\bibnamefont
  {Horodecki}}, \bibinfo {author} {\bibfnamefont {P.}~\bibnamefont
  {Horodecki}}, \bibinfo {author} {\bibfnamefont {M.}~\bibnamefont
  {Horodecki}}, \ and\ \bibinfo {author} {\bibfnamefont {K.}~\bibnamefont
  {Horodecki}},\ }\href@noop {} {\bibfield  {journal} {\bibinfo  {journal}
  {Rev.~Mod.~Phys.}\ }\textbf {\bibinfo {volume} {81}},\ \bibinfo {pages} {865}
  (\bibinfo {year} {2009})}\BibitemShut {NoStop}%
\bibitem [{\citenamefont {Modi}\ \emph {et~al.}(2012)\citenamefont {Modi},
  \citenamefont {Brodutch}, \citenamefont {Cable}, \citenamefont {Paterek},\
  and\ \citenamefont {Vedral}}]{modi2012classical}%
  \BibitemOpen
  \bibfield  {author} {\bibinfo {author} {\bibfnamefont {K.}~\bibnamefont
  {Modi}}, \bibinfo {author} {\bibfnamefont {A.}~\bibnamefont {Brodutch}},
  \bibinfo {author} {\bibfnamefont {H.}~\bibnamefont {Cable}}, \bibinfo
  {author} {\bibfnamefont {T.}~\bibnamefont {Paterek}}, \ and\ \bibinfo
  {author} {\bibfnamefont {V.}~\bibnamefont {Vedral}},\ }\href@noop {}
  {\bibfield  {journal} {\bibinfo  {journal} {Reviews of Modern Physics}\
  }\textbf {\bibinfo {volume} {84}},\ \bibinfo {pages} {1655} (\bibinfo {year}
  {2012})}\BibitemShut {NoStop}%
\bibitem [{\citenamefont {Nielsen}\ and\ \citenamefont
  {Chuang}(2010)}]{nielsen2010quantum}%
  \BibitemOpen
  \bibfield  {author} {\bibinfo {author} {\bibfnamefont {M.~A.}\ \bibnamefont
  {Nielsen}}\ and\ \bibinfo {author} {\bibfnamefont {I.~L.}\ \bibnamefont
  {Chuang}},\ }\href@noop {} {\emph {\bibinfo {title} {Quantum computation and
  quantum information}}}\ (\bibinfo  {publisher} {Cambridge university press},\
  \bibinfo {year} {2010})\BibitemShut {NoStop}%
\bibitem [{\citenamefont {Zurek}(2003)}]{zurek2003quantum}%
  \BibitemOpen
  \bibfield  {author} {\bibinfo {author} {\bibfnamefont {W.~H.}\ \bibnamefont
  {Zurek}},\ }\href@noop {} {\bibfield  {journal} {\bibinfo  {journal}
  {Physical Review A}\ }\textbf {\bibinfo {volume} {67}},\ \bibinfo {pages}
  {012320} (\bibinfo {year} {2003})}\BibitemShut {NoStop}%
\bibitem [{\citenamefont {Henderson}\ and\ \citenamefont
  {Vedral}(2001)}]{Henderson:JournalOfPhysicsAMathematicalAndGeneral:2001}%
  \BibitemOpen
  \bibfield  {author} {\bibinfo {author} {\bibfnamefont {L.}~\bibnamefont
  {Henderson}}\ and\ \bibinfo {author} {\bibfnamefont {V.}~\bibnamefont
  {Vedral}},\ }\href {\doibase 10.1088/0305-4470/34/35/315} {\bibfield
  {journal} {\bibinfo  {journal} {J. Phys. A:Math Gen}\ }\textbf {\bibinfo
  {volume} {34}},\ \bibinfo {pages} {6899} (\bibinfo {year}
  {2001})}\BibitemShut {NoStop}%
\bibitem [{\citenamefont {Luo}(2008{\natexlab{a}})}]{luo08a}%
  \BibitemOpen
  \bibfield  {author} {\bibinfo {author} {\bibfnamefont {S.}~\bibnamefont
  {Luo}},\ }\href@noop {} {\bibfield  {journal} {\bibinfo  {journal} {Physical
  Review~A}\ }\textbf {\bibinfo {volume} {77}},\ \bibinfo {pages} {022301}
  (\bibinfo {year} {2008}{\natexlab{a}})}\BibitemShut {NoStop}%
\bibitem [{\citenamefont {Wu}\ \emph {et~al.}(2009)\citenamefont {Wu},
  \citenamefont {Poulsen},\ and\ \citenamefont {M{\o}lmer}}]{wu09a}%
  \BibitemOpen
  \bibfield  {author} {\bibinfo {author} {\bibfnamefont {S.}~\bibnamefont
  {Wu}}, \bibinfo {author} {\bibfnamefont {U.~V.}\ \bibnamefont {Poulsen}}, \
  and\ \bibinfo {author} {\bibfnamefont {K.}~\bibnamefont {M{\o}lmer}},\
  }\href@noop {} {\bibfield  {journal} {\bibinfo  {journal} {Physical
  Review~A}\ }\textbf {\bibinfo {volume} {80}},\ \bibinfo {pages} {032319}
  (\bibinfo {year} {2009})}\BibitemShut {NoStop}%
\bibitem [{\citenamefont {Lang}\ \emph {et~al.}(2011)\citenamefont {Lang},
  \citenamefont {Caves},\ and\ \citenamefont
  {Shaji}}]{LANG:IntJQuanumInform:2011}%
  \BibitemOpen
  \bibfield  {author} {\bibinfo {author} {\bibfnamefont {M.~D.}\ \bibnamefont
  {Lang}}, \bibinfo {author} {\bibfnamefont {C.~M.}\ \bibnamefont {Caves}}, \
  and\ \bibinfo {author} {\bibfnamefont {A.}~\bibnamefont {Shaji}},\ }\href
  {\doibase 10.1142/s021974991100826x} {\bibfield  {journal} {\bibinfo
  {journal} {Int. J. Quanum Inform.}\ }\textbf {\bibinfo {volume} {09}},\
  \bibinfo {pages} {1553} (\bibinfo {year} {2011})}\BibitemShut {NoStop}%
\bibitem [{\citenamefont {Singh}\ and\ \citenamefont
  {Pati}(2014)}]{singh2014quantum}%
  \BibitemOpen
  \bibfield  {author} {\bibinfo {author} {\bibfnamefont {U.}~\bibnamefont
  {Singh}}\ and\ \bibinfo {author} {\bibfnamefont {A.~K.}\ \bibnamefont
  {Pati}},\ }\href@noop {} {\bibfield  {journal} {\bibinfo  {journal} {Annals
  of Physics}\ }\textbf {\bibinfo {volume} {343}},\ \bibinfo {pages} {141}
  (\bibinfo {year} {2014})}\BibitemShut {NoStop}%
\bibitem [{\citenamefont {Shaji}\ \emph {et~al.}(2015)\citenamefont {Shaji},
  \citenamefont {Shaji} \emph {et~al.}}]{shaji2015weak}%
  \BibitemOpen
  \bibfield  {author} {\bibinfo {author} {\bibfnamefont {N.}~\bibnamefont
  {Shaji}}, \bibinfo {author} {\bibfnamefont {A.}~\bibnamefont {Shaji}},  \emph
  {et~al.},\ }\href@noop {} {\bibfield  {journal} {\bibinfo  {journal} {arXiv
  preprint arXiv:1511.09224}\ } (\bibinfo {year} {2015})}\BibitemShut {NoStop}%
\bibitem [{\citenamefont {Aharonov}\ \emph {et~al.}(1988)\citenamefont
  {Aharonov}, \citenamefont {Albert},\ and\ \citenamefont
  {Vaidman}}]{aharonov1988result}%
  \BibitemOpen
  \bibfield  {author} {\bibinfo {author} {\bibfnamefont {Y.}~\bibnamefont
  {Aharonov}}, \bibinfo {author} {\bibfnamefont {D.~Z.}\ \bibnamefont
  {Albert}}, \ and\ \bibinfo {author} {\bibfnamefont {L.}~\bibnamefont
  {Vaidman}},\ }\href@noop {} {\bibfield  {journal} {\bibinfo  {journal}
  {Physical review letters}\ }\textbf {\bibinfo {volume} {60}},\ \bibinfo
  {pages} {1351} (\bibinfo {year} {1988})}\BibitemShut {NoStop}%
\bibitem [{\citenamefont {Duck}\ \emph {et~al.}(1989)\citenamefont {Duck},
  \citenamefont {Stevenson},\ and\ \citenamefont {Sudarshan}}]{duck1989sense}%
  \BibitemOpen
  \bibfield  {author} {\bibinfo {author} {\bibfnamefont {I.}~\bibnamefont
  {Duck}}, \bibinfo {author} {\bibfnamefont {P.}~\bibnamefont {Stevenson}}, \
  and\ \bibinfo {author} {\bibfnamefont {E.}~\bibnamefont {Sudarshan}},\
  }\href@noop {} {\bibfield  {journal} {\bibinfo  {journal} {Physical Review
  D}\ }\textbf {\bibinfo {volume} {40}},\ \bibinfo {pages} {2112} (\bibinfo
  {year} {1989})}\BibitemShut {NoStop}%
\bibitem [{\citenamefont {Dixon}\ \emph {et~al.}(2009)\citenamefont {Dixon},
  \citenamefont {Starling}, \citenamefont {Jordan},\ and\ \citenamefont
  {Howell}}]{dixon2009ultrasensitive}%
  \BibitemOpen
  \bibfield  {author} {\bibinfo {author} {\bibfnamefont {P.~B.}\ \bibnamefont
  {Dixon}}, \bibinfo {author} {\bibfnamefont {D.~J.}\ \bibnamefont {Starling}},
  \bibinfo {author} {\bibfnamefont {A.~N.}\ \bibnamefont {Jordan}}, \ and\
  \bibinfo {author} {\bibfnamefont {J.~C.}\ \bibnamefont {Howell}},\
  }\href@noop {} {\bibfield  {journal} {\bibinfo  {journal} {Physical review
  letters}\ }\textbf {\bibinfo {volume} {102}},\ \bibinfo {pages} {173601}
  (\bibinfo {year} {2009})}\BibitemShut {NoStop}%
\bibitem [{\citenamefont {Hosten}\ and\ \citenamefont
  {Kwiat}(2008)}]{hosten2008observation}%
  \BibitemOpen
  \bibfield  {author} {\bibinfo {author} {\bibfnamefont {O.}~\bibnamefont
  {Hosten}}\ and\ \bibinfo {author} {\bibfnamefont {P.}~\bibnamefont {Kwiat}},\
  }\href@noop {} {\bibfield  {journal} {\bibinfo  {journal} {Science}\ }\textbf
  {\bibinfo {volume} {319}},\ \bibinfo {pages} {787} (\bibinfo {year}
  {2008})}\BibitemShut {NoStop}%
\bibitem [{\citenamefont {P{\"u}tz}\ \emph {et~al.}(2016)\citenamefont
  {P{\"u}tz}, \citenamefont {Barnea}, \citenamefont {Gisin},\ and\
  \citenamefont {Martin}}]{Putz:2016tm}%
  \BibitemOpen
  \bibfield  {author} {\bibinfo {author} {\bibfnamefont {G.}~\bibnamefont
  {P{\"u}tz}}, \bibinfo {author} {\bibfnamefont {T.}~\bibnamefont {Barnea}},
  \bibinfo {author} {\bibfnamefont {N.}~\bibnamefont {Gisin}}, \ and\ \bibinfo
  {author} {\bibfnamefont {A.}~\bibnamefont {Martin}},\ }\href@noop {} {\
  (\bibinfo {year} {2016})},\ \Eprint {http://arxiv.org/abs/arXiv: 1610.04464}
  {arXiv: 1610.04464} \BibitemShut {NoStop}%
\bibitem [{\citenamefont {Oreshkov}\ and\ \citenamefont
  {Brun}(2005)}]{oreshkov2005weak}%
  \BibitemOpen
  \bibfield  {author} {\bibinfo {author} {\bibfnamefont {O.}~\bibnamefont
  {Oreshkov}}\ and\ \bibinfo {author} {\bibfnamefont {T.~A.}\ \bibnamefont
  {Brun}},\ }\href@noop {} {\bibfield  {journal} {\bibinfo  {journal} {Physical
  review letters}\ }\textbf {\bibinfo {volume} {95}},\ \bibinfo {pages}
  {110409} (\bibinfo {year} {2005})}\BibitemShut {NoStop}%
\bibitem [{\citenamefont {Dressel}\ and\ \citenamefont
  {Jordan}(2012)}]{dressel2012significance}%
  \BibitemOpen
  \bibfield  {author} {\bibinfo {author} {\bibfnamefont {J.}~\bibnamefont
  {Dressel}}\ and\ \bibinfo {author} {\bibfnamefont {A.}~\bibnamefont
  {Jordan}},\ }\href@noop {} {\bibfield  {journal} {\bibinfo  {journal}
  {Physical Review A}\ }\textbf {\bibinfo {volume} {85}},\ \bibinfo {pages}
  {012107} (\bibinfo {year} {2012})}\BibitemShut {NoStop}%
\bibitem [{\citenamefont {Knee}\ \emph {et~al.}(2016)\citenamefont {Knee},
  \citenamefont {Combes}, \citenamefont {Ferrie},\ and\ \citenamefont
  {Gauger}}]{knee2016weak}%
  \BibitemOpen
  \bibfield  {author} {\bibinfo {author} {\bibfnamefont {G.~C.}\ \bibnamefont
  {Knee}}, \bibinfo {author} {\bibfnamefont {J.}~\bibnamefont {Combes}},
  \bibinfo {author} {\bibfnamefont {C.}~\bibnamefont {Ferrie}}, \ and\ \bibinfo
  {author} {\bibfnamefont {E.~M.}\ \bibnamefont {Gauger}},\ }\href@noop {}
  {\bibfield  {journal} {\bibinfo  {journal} {Quantum Measurements and Quantum
  Metrology}\ }\textbf {\bibinfo {volume} {3}},\ \bibinfo {pages} {32}
  (\bibinfo {year} {2016})}\BibitemShut {NoStop}%
\bibitem [{\citenamefont {Guo}\ and\ \citenamefont
  {Wu}(2014)}]{guo2014quantum}%
  \BibitemOpen
  \bibfield  {author} {\bibinfo {author} {\bibfnamefont {Y.}~\bibnamefont
  {Guo}}\ and\ \bibinfo {author} {\bibfnamefont {S.}~\bibnamefont {Wu}},\
  }\href@noop {} {\bibfield  {journal} {\bibinfo  {journal} {Scientific
  Reports}\ }\textbf {\bibinfo {volume} {4}},\ \bibinfo {pages} {7179}
  (\bibinfo {year} {2014})}\BibitemShut {NoStop}%
\bibitem [{\citenamefont {Luo}(2008{\natexlab{b}})}]{luo2008quantum}%
  \BibitemOpen
  \bibfield  {author} {\bibinfo {author} {\bibfnamefont {S.}~\bibnamefont
  {Luo}},\ }\href@noop {} {\bibfield  {journal} {\bibinfo  {journal} {Physical
  Review A}\ }\textbf {\bibinfo {volume} {77}},\ \bibinfo {pages} {042303}
  (\bibinfo {year} {2008}{\natexlab{b}})}\BibitemShut {NoStop}%
\bibitem [{\citenamefont {Vinjanampathy}\ and\ \citenamefont
  {Rau}(2011)}]{vinjanampathy2011calculation}%
  \BibitemOpen
  \bibfield  {author} {\bibinfo {author} {\bibfnamefont {S.}~\bibnamefont
  {Vinjanampathy}}\ and\ \bibinfo {author} {\bibfnamefont {A.}~\bibnamefont
  {Rau}},\ }\href@noop {} {\bibfield  {journal} {\bibinfo  {journal} {arXiv
  preprint arXiv:1106.4488}\ }\textbf {\bibinfo {volume} {88}} (\bibinfo {year}
  {2011})}\BibitemShut {NoStop}%
\bibitem [{\citenamefont {Yurischev}(2015)}]{yurischev2015quantum}%
  \BibitemOpen
  \bibfield  {author} {\bibinfo {author} {\bibfnamefont {M.}~\bibnamefont
  {Yurischev}},\ }\href@noop {} {\bibfield  {journal} {\bibinfo  {journal}
  {Quantum Information Processing}\ }\textbf {\bibinfo {volume} {14}},\
  \bibinfo {pages} {3399} (\bibinfo {year} {2015})}\BibitemShut {NoStop}%
\bibitem [{\citenamefont {Buscemi}\ and\ \citenamefont
  {Bordone}(2013)}]{buscemi2013time}%
  \BibitemOpen
  \bibfield  {author} {\bibinfo {author} {\bibfnamefont {F.}~\bibnamefont
  {Buscemi}}\ and\ \bibinfo {author} {\bibfnamefont {P.}~\bibnamefont
  {Bordone}},\ }\href@noop {} {\bibfield  {journal} {\bibinfo  {journal}
  {Physical Review A}\ }\textbf {\bibinfo {volume} {87}},\ \bibinfo {pages}
  {042310} (\bibinfo {year} {2013})}\BibitemShut {NoStop}%
\bibitem [{\citenamefont {Guo}\ \emph {et~al.}(2016)\citenamefont {Guo},
  \citenamefont {Zeng},\ and\ \citenamefont {Wang}}]{guo2016pairwise}%
  \BibitemOpen
  \bibfield  {author} {\bibinfo {author} {\bibfnamefont {Y.-N.}\ \bibnamefont
  {Guo}}, \bibinfo {author} {\bibfnamefont {K.}~\bibnamefont {Zeng}}, \ and\
  \bibinfo {author} {\bibfnamefont {G.-Y.}\ \bibnamefont {Wang}},\ }\href@noop
  {} {\bibfield  {journal} {\bibinfo  {journal} {International Journal of
  Theoretical Physics}\ }\textbf {\bibinfo {volume} {55}},\ \bibinfo {pages}
  {2894} (\bibinfo {year} {2016})}\BibitemShut {NoStop}%
\bibitem [{\citenamefont {Rulli}\ and\ \citenamefont
  {Sarandy}(2011)}]{rulli2011global}%
  \BibitemOpen
  \bibfield  {author} {\bibinfo {author} {\bibfnamefont {C.}~\bibnamefont
  {Rulli}}\ and\ \bibinfo {author} {\bibfnamefont {M.}~\bibnamefont
  {Sarandy}},\ }\href@noop {} {\bibfield  {journal} {\bibinfo  {journal}
  {Physical Review A}\ }\textbf {\bibinfo {volume} {84}},\ \bibinfo {pages}
  {042109} (\bibinfo {year} {2011})}\BibitemShut {NoStop}%
\bibitem [{\citenamefont {Madhok}\ and\ \citenamefont
  {Datta}(2011)}]{madhok2011interpreting}%
  \BibitemOpen
  \bibfield  {author} {\bibinfo {author} {\bibfnamefont {V.}~\bibnamefont
  {Madhok}}\ and\ \bibinfo {author} {\bibfnamefont {A.}~\bibnamefont {Datta}},\
  }\href@noop {} {\bibfield  {journal} {\bibinfo  {journal} {Physical Review
  A}\ }\textbf {\bibinfo {volume} {83}},\ \bibinfo {pages} {032323} (\bibinfo
  {year} {2011})}\BibitemShut {NoStop}%
\bibitem [{\citenamefont {Cavalcanti}\ \emph {et~al.}(2011)\citenamefont
  {Cavalcanti}, \citenamefont {Aolita}, \citenamefont {Boixo}, \citenamefont
  {Modi}, \citenamefont {Piani},\ and\ \citenamefont
  {Winter}}]{cavalcanti2011operational}%
  \BibitemOpen
  \bibfield  {author} {\bibinfo {author} {\bibfnamefont {D.}~\bibnamefont
  {Cavalcanti}}, \bibinfo {author} {\bibfnamefont {L.}~\bibnamefont {Aolita}},
  \bibinfo {author} {\bibfnamefont {S.}~\bibnamefont {Boixo}}, \bibinfo
  {author} {\bibfnamefont {K.}~\bibnamefont {Modi}}, \bibinfo {author}
  {\bibfnamefont {M.}~\bibnamefont {Piani}}, \ and\ \bibinfo {author}
  {\bibfnamefont {A.}~\bibnamefont {Winter}},\ }\href@noop {} {\bibfield
  {journal} {\bibinfo  {journal} {Physical Review A}\ }\textbf {\bibinfo
  {volume} {83}},\ \bibinfo {pages} {032324} (\bibinfo {year}
  {2011})}\BibitemShut {NoStop}%
\bibitem [{\citenamefont {Daki{\'c}}\ \emph {et~al.}(2012)\citenamefont
  {Daki{\'c}}, \citenamefont {Lipp}, \citenamefont {Ma}, \citenamefont
  {Ringbauer}, \citenamefont {Kropatschek}, \citenamefont {Barz}, \citenamefont
  {Paterek}, \citenamefont {Vedral}, \citenamefont {Zeilinger}, \citenamefont
  {Brukner} \emph {et~al.}}]{dakic2012quantum}%
  \BibitemOpen
  \bibfield  {author} {\bibinfo {author} {\bibfnamefont {B.}~\bibnamefont
  {Daki{\'c}}}, \bibinfo {author} {\bibfnamefont {Y.~O.}\ \bibnamefont {Lipp}},
  \bibinfo {author} {\bibfnamefont {X.}~\bibnamefont {Ma}}, \bibinfo {author}
  {\bibfnamefont {M.}~\bibnamefont {Ringbauer}}, \bibinfo {author}
  {\bibfnamefont {S.}~\bibnamefont {Kropatschek}}, \bibinfo {author}
  {\bibfnamefont {S.}~\bibnamefont {Barz}}, \bibinfo {author} {\bibfnamefont
  {T.}~\bibnamefont {Paterek}}, \bibinfo {author} {\bibfnamefont
  {V.}~\bibnamefont {Vedral}}, \bibinfo {author} {\bibfnamefont
  {A.}~\bibnamefont {Zeilinger}}, \bibinfo {author} {\bibfnamefont
  {{\v{C}}.}~\bibnamefont {Brukner}},  \emph {et~al.},\ }\href@noop {}
  {\bibfield  {journal} {\bibinfo  {journal} {Nature Physics}\ }\textbf
  {\bibinfo {volume} {8}},\ \bibinfo {pages} {666} (\bibinfo {year}
  {2012})}\BibitemShut {NoStop}%
\bibitem [{\citenamefont {Pirandola}(2014)}]{pirandola2014quantum}%
  \BibitemOpen
  \bibfield  {author} {\bibinfo {author} {\bibfnamefont {S.}~\bibnamefont
  {Pirandola}},\ }\href@noop {} {\bibfield  {journal} {\bibinfo  {journal}
  {Scientific Reports}\ }\textbf {\bibinfo {volume} {4}},\ \bibinfo {pages}
  {6956} (\bibinfo {year} {2014})}\BibitemShut {NoStop}%
\end{thebibliography}%

\end{document}